# IONOSPHERIC DISPERSION COMPENSATION USING A NOVEL MICROWAVE DE-DISPERSION FILTER

**Paul Roberts**

*CSIRO, Australia Telescope National Facility,*

*PO Box 76, Epping, Australia, 2122*

*Paul.Roberts@csiro.au*

**ABSTRACT**

Free electrons in the ionosphere lead to significant group delay dispersion for signals in the megahertz and low gigahertz range. A novel microwave filter is presented that is capable of compensating for the non-linear ionospheric dispersion over a 600 MHz bandwidth between 1.2-1.8 GHz. The design method is general and is not limited to this particular frequency range but can provide an arbitrary phase and amplitude response over any frequency range. Several of the filters were constructed and used in an experiment to detect short radio pulse emission from the lunar regolith.

**INTRODUCTION**

The ionosphere is a dispersive medium so that radio waves propagating through the ionosphere suffer a frequency dependent delay. This delay is given to a very high level of accuracy by [1]

$$\tau_{gd}(f) = \frac{40.3\,TEC}{f^2} \tag{1}$$

where TEC, the total electron content, is the line integral of the free electron density along the path of propagation and f is the frequency of the propagating wave. TEC values vary widely over a diurnal cycle and also as a function of solar activity linked to the approximately 11 year solar cycle. Lowest values of TEC, and thus delay, occur during the hours of local midnight to sometime before dawn.

The ionospheric group velocity dispersion can have especially serious effects for short pulse signals as the dispersion can smear the pulses out to the point where they are below the instrumental detection threshold. The motivation for the current work was a search for pulses of order 1 ns duration over the frequency range 1.2-1.8 GHz. At the current point in the solar cycle, and during local night time, signals across this band suffer approximately 5 ns of dispersion so that without compensation, detection sensitivity would be severely compromised.

**NON-UNIFORM TRANSMISSION LINE FILTERS**

In order to compensate for the ionospheric delay a microwave filter is required that has group delay of opposite sign to that of the ionosphere and with minimal amplitude distortion. Thus the ideal is an all pass filter with prescribed phase response. Microwave filters with arbitrary phase response cannot be easily designed by standard filter design methods which generally give complex frequency responses defined by particular prototype polynomials (Bessel, Chebyshev etc). An exception is the class of non-uniform transmission line filters [2]. In the microstrip implementation this filter takes the form of a microstrip line with continuously varying width. The local changes in width, and hence transmission line impedance, lead to frequency dependent reflection and transmission of the propagating signal at each point along the transmission line. By taking account of the multiple reflections within the structure it is possible to design a filter with arbitrary complex reflection coefficient at the input. The inverse scattering method is employed to design these filters. Either a direct solution method involving solving of coupled integro-differential equations [2], or an iterative "layer stripping" method [3], can be employed to solve the inverse scattering problem. In the present design the layer stripping method is used.

**DESIGN OF COMPENSATING FILTER**

The delay introduced by the ionosphere is as given by equation 1. In order to compensate this a filter is required with group delay response given by

$$\tau(f) = C - \frac{40.3\,TEC}{f^2} \qquad (2)$$

Here C is an arbitrary delay offset. In practice it is chosen such that the group delay of the filter goes to zero just below the required lower band edge. This prevents the filter design algorithm from synthesising a long section of uniform transmission line at the beginning of the filter and ensures a realisable (positive) group delay. Given this group delay the required phase response of the filter is

$$\phi(f) = -(2\pi C f + \frac{40.3\,TEC}{2\pi f}). \qquad (3)$$

Ideally the amplitude response is unity. In practice the amplitude response must be band limited in order for the filter to be realisable. To ensure realisability, in the current design the all pass response is multiplied by the amplitude of a $4^{th}$ order Bessel bandpass polynomial encompassing the design range of 1.2 – 1.8 GHz. Amplitude selectivity is not of concern for this delay compensating filter so a high order / steep cut-off amplitude function was not required. The present filter was designed to correct 6.8 ns of dispersion across the 1.2 – 1.8 GHz.

The design proceeds by sampling the complex frequency response, defined by the phase and amplitude responses, at 256 points between 0 and 4 GHz. The discrete inverse Fourier transform of this is calculated. This gives the real valued discrete impulse response of the desired filter. The problem now is to calculate the impedance profile of the transmission line filter that results in the given impulse response for a wave incident upon the filter. This is the inverse scattering problem. Given that the impulse response is discrete the iterative layer stripping algorithm can be used [3]. The solution corresponds to a transmission line filter composed of stepped impedance sections. As the number of sample points increases beyond the 256 chosen here the resulting transmission line filter approaches a continuous smooth profile.

The layer stripping algorithm functions by using the causal nature of the impulse response. The first impulse response sample can only be a result of reflection at the first impedance step of the transmission line filter. This allows determination of the impedance of the first element of the filter. The effect of the first element can then be removed from the propagating impulse to calculate the next impedance step and so forth "stripping" the layers back through all elements of the filter.

Given the impedance profile along the filter the corresponding width of the transmission line can be found using standard design formulas for the chosen transmission line geometry of the application [4].

The layer stripping procedure generally results in a very long filter. After some point, further changes in impedance have negligible effect so the filter can be truncated. At the truncation point a load equal to the line impedance is placed to terminate the filter. The filter may also be prematurely truncated in order to more aggressively decrease the length of the filter at the price of increased group delay ripple and amplitude distortion. The expected frequency response of the truncated filter can be simulated using cascaded transmission matrices representing the uniform segments of cascaded transmission line. This is most easily accomplished using standard microwave design software packages.

**RESULTS**

A filter with total dispersion compensation of 6.8 ns across 1.2 – 1.8 GHz was designed. The resulting filter was strongly truncated in order to reduce the filter length. The total length of the filter to correct for the 6.8 ns of dispersion was 9.375 ns. The filter was fabricated on Rogers 5880 substrate (permittivity = 2.2) of 31 thou thickness. The total filter length on this substrate is 2.64 m. In order to minimise the area required for the filter it was curled up into a spiral structure of diameter 25 cm. The filter was fabricated in house using standard PCB fabrication techniques. Figure 1. below shows the mask pattern for the completed filter.

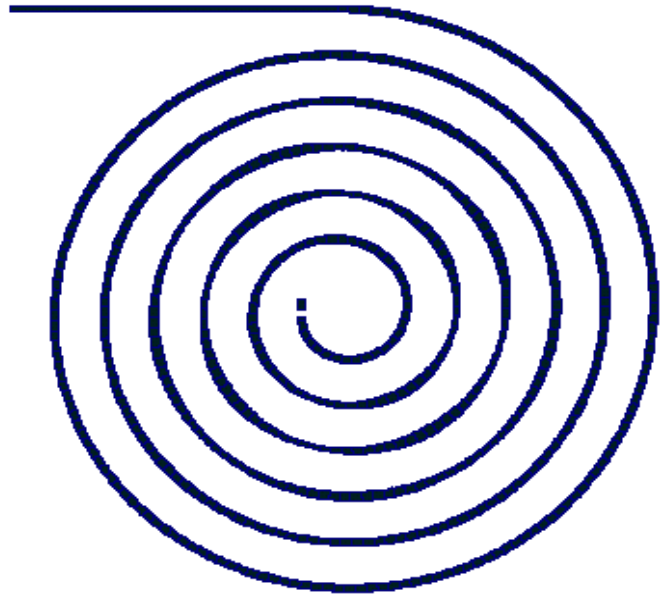

Figure 1. Mask pattern of completed filter.

The amplitude and group delay response of the filter were measured after attaching an SMA microstrip launch connector at the input of the filter. Figure 2 below shows the measured results compared to simulation of the truncated filter and the ideal design curve.

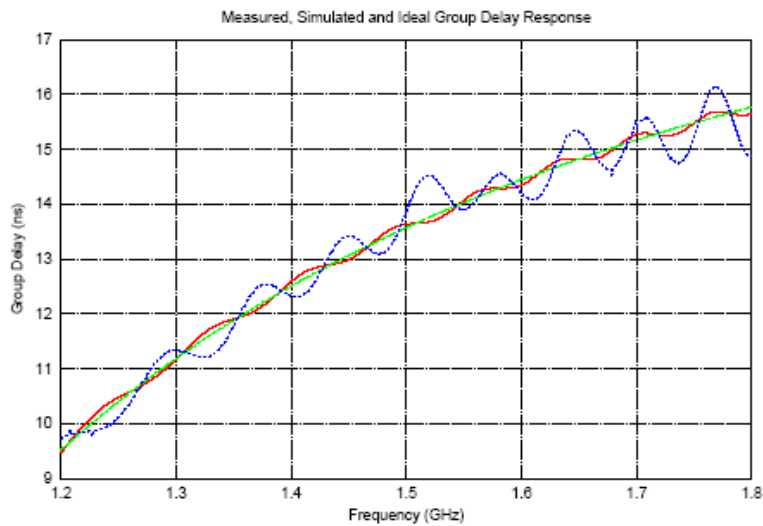

Figure 2. Measured group delay (blue) vs simulated (red) and ideal (green).

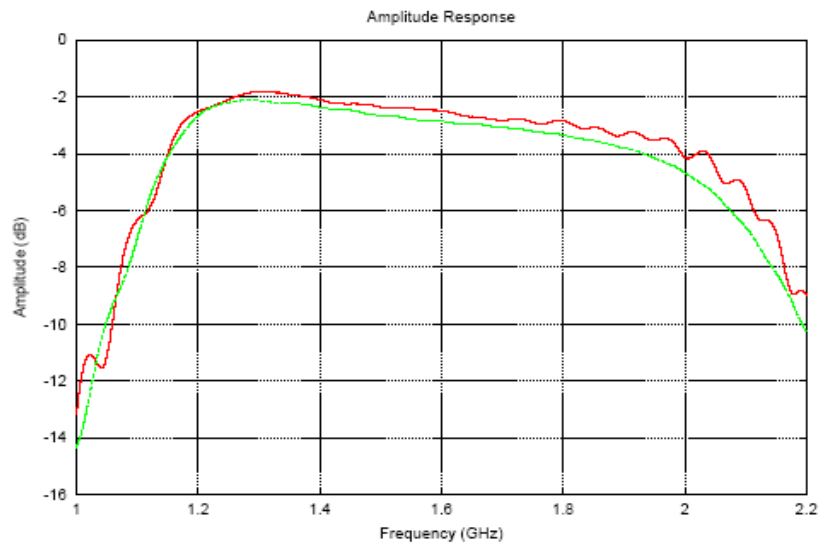

Figure 3. Measured amplitude response (red) vs simulated (green).

The most significant departure from simulation is the additional group delay ripple seen in the group delay response. This was determined to be due to mismatch at the interface of the microstrip line and microstrip launch connector. On the selected substrate the width of the 50 ohm microstrip input line is 1.58 mm. The width of the centre pin of the launch connector is 0.3 mm leading to a significant impedance discontinuity and subsequent reflection. This interface should be improved in subsequent designs. Loss across the band is approximately 1.5 dB caused by ohmic and dielectric losses in the microstrip. This could have been corrected for in the design algorithm by using an inverse weighted amplitude response but in the particular application for which this filter was designed the slope was beneficial, as the signal being searched for has a decreasing signal to noise ratio with frequency, so the slope gives a more optimal weighting to the higher signal to noise ratio portions of the signal.

**CONCLUSION**

A microwave filter to correct for ionospheric dispersion over 1.2-1.8 GHz was designed using an inverse scattering algorithm. The filter was fabricated in microstrip and compared to theoretical and simulated predictions. The need for an improved connector interface was highlighted for subsequent designs. Six of the filters were deployed on the Australia Telescope Compact Array for an experiment to search for short pulse emission from the lunar regolith in June 2007.

**REFERENCES**

[1] J. A. Klobuchar. "Ionospheric effects on GPS" in Global Positioning System: Theory and Applications, Parkinson, B. W. and Spilker Jr, J. J., eds., Vol. 1, vol. 163 of Progess in Astronautics and Aeronautics. American Institute of Aeronautics and Astronautics, Washington, U.S.A., 1996.

[2] P. Roberts and G. Town, "Design of microwave filters by inverse scattering", *IEEE Trans. Microwave Theory and Techniques*, vol. 43, pp. 739-743, 1995.

[3] A. M. Bruckstein and T. Kailath, "Inverse scattering for discrete transmission-line models," *SIAM Review*, vol. 29, no. 3, pp. 359-389, 1987.

[4] E. H. Fooks and R. A. Zakarevicius, *Microwave Engineering using Microstrip Circuits*, Prentice-Hall, 1990